# InAs quantum dot in a needlelike tapered InP nanowire: a telecom band single photon source monolithically grown on silicon

Ali Jaffal,*[1a,b] Walid Redjem,[2] Philippe Regreny,[1b] Hai Son Nguyen,[1b] Sébastien Cueff,[1b] Xavier Letartre,[1b] Gilles Patriarche,[3] Emmanuel Rousseau,[2] Guillaume Cassabois,[2] Michel Gendry[1b] and Nicolas Chauvin[1a]

[1]Université de Lyon, Institut des Nanotechnologies de Lyon, UMR 5270 CNRS,

[a]INSA de Lyon, 7 avenue Jean Capelle, 69621 Villeurbanne cedex, France

[b]Ecole Centrale de Lyon, 36 avenue Guy de Collongue, 69134 Ecully cedex, France.

[2]Université de Montpellier, Laboratoire Charles Coulomb, UMR 5221 CNRS, Place Eugène Bataillon, F-34095 Montpellier Cedex 5, France

[3]Université Paris-Saclay, Centre de Nanosciences et de Nanotechnologies, UMR 9001 CNRS,

10 boulevard Thomas Gobert, 91120 Palaiseau, France

Email : ali.jaffal@insa-lyon.fr

Realizing single photon sources emitting in the telecom band on silicon substrates is essential to reach complementary-metal-oxide-semiconductor (CMOS) compatible devices that secure communications over long distances. In this work, we propose the monolithic growth of needlelike tapered InAs/InP quantum dot-nanowires (QD-NWs) on silicon substrates with a small taper angle and a nanowire diameter tailored to support a single mode waveguide. Such a NW geometry is obtained by a controlled balance over axial and radial growths during the gold-catalyzed growth of the NWs by molecular beam epitaxy. This allows us to investigate the impact of the taper angle on the emission properties of a single InAs/InP QD-NW. At room temperature, a Gaussian far-field emission profile in the telecom O-band with a beam divergence angle $\theta = 30°$ is demonstrated from a single InAs QD embedded in a 2° tapered InP NW. Moreover, single photon emission is observed at cryogenic temperature for an off-resonant excitation and the best result, $g^2(0) = 0.05$, is obtained for a 7° tapered NW. This all-encompassing study paves the way for the monolithic growth on silicon of an efficient single photon source in the telecom band based on InAs/InP QD-NWs.

Keywords: Tapered nanowire, Quantum dot, InAs/InP, Vapor-Liquid-Solid Molecular Beam Epitaxy, Gaussian far-field emission profile, Telecom band, Silicon integration

Non-classical light sources emitting at optical communication wavelength bands are of prime importance to quantum communication applications. One of these sources is the single photon source (SPS) which is the building block for realizing scalable on-chip devices for quantum information processing. Once the single photons are generated, it is then required to manipulate these photons either to encode the information or to make the measurements. This means that the SPS must be integrated with compact photonic devices that can combine many optical components. Thanks to its large refractive index, silicon (Si) facilitates the fitting of a high number of optical components into a small device size making it a powerful platform for photonic integrated circuits.[1] Moreover, Si appears to be the most compatible material, due to the maturity of the complementary-metal-oxide-semiconductor (CMOS) fabrication methods, to combine electronics with photonics.[2] However, Si is an indirect band gap material which makes it a very poor light source. This issue can be solved by the monolithic growth, bonding or pick-and-place procedure of III-V materials on Si to fabricate CMOS compatible SPS.[3–6] In addition to the substrate choice, SPS emitting in the 1.3 µm and 1.5 µm telecom windows are required to reduce the losses for fiber-based long-haul communications. In particular, InAs quantum dots (QDs) have proven to be good candidates as efficient non-classical light sources in these bands.[7,8] This requires embedding the QD in a nanophotonic structure such as nanowires (NWs),[9] micropillars,[10] optical horns,[11] photonic crystal cavities[12,13] and waveguides[14] in order to guide and efficiently extract the light in free space.[15]

The monolithic growth of InAs/InP QD-NWs on Si can be realized using the vapor-liquid-solid (VLS) growth method which ensures the epitaxial relationship between the NWs and the substrate.[16–20] With this method, InAs/InP QD-NW SPS emitting in telecom band can be achieved. However, the NW diameter must be optimized to couple the QD emission to the NW fundamental waveguide mode,[9] $HE_{11}$, and the tip of the NW has to be tailored to direct the single photons towards the collection optics with a less diverging output beam.[21] Tailoring the upper part of a photonic wire has recently shown to be an advanced solution to funnel the emission of a point-like emitter into a Gaussian free-space beam[22] by fabricating giant "photonic trumpets" using a top down approach.[23] Another strategy is to fabricate NWs with a narrow-tapered "needlelike" tip using a bottom up or a top-down approach. The effect of the NW taper on the light emitted by the QD has been demonstrated theoretically and experimentally: a taper angle, α, smaller than 5° must be achieved to increase the photon extraction from the NW[24] and therefore to enhance the source brightness.[25] In this case, the light reflection and scattering at the NW top facet-air interface will be minimized allowing the guided light to couple out from the NW to free space. A Gaussian far-field emission at 980 nm from a tapered QD-NW grown on InP substrate was obtained due to the coupling between the $HE_{11}$ mode and the QD emission allowing efficient light collection and coupling to a single mode fiber.[26] To this day, most of the works done in literature concerning the far-field emission of single QD-NW or NW, is either in the visible or in the near-infrared 900-1000 nm range.[22,26–29] However, no study has been reported yet on a single photon emission with a Gaussian far-field radiation in the telecom band from single QD-NWs monolithically grown on Si substrates.

In this letter, we experimentally demonstrate a Gaussian emission profile in the telecom O-band from a single InAs/InP QD-NW monolithically grown on Si(111) substrates by solid-source molecular beam epitaxy (ss-MBE) using the VLS method. The control on the NW shape allowed us to record an estimated source efficiency, $\epsilon(\theta) \approx 42\%$, as well as detecting a single photon emission from monolithic InAs/InP QD-NWs on Si.

The InAs/InP QD-NWs were grown on Si(111) substrates by VLS-MBE using In-Au droplets in-situ deposited at 500°C as catalyst (see methods).[30] To study a single QD emission we have chosen to grow a low density of InAs/InP QD-NWs on Si. This was achieved with Au-In droplets by evaporating small amounts of In and Au on the Si substrate with In and Au beam equivalent pressures (BEP) equal to $2.5 \times 10^{-7}$ torr and $4.2 \times 10^{-9}$ torr, respectively, corresponding to an In/Au BEP ratio ≈ 60. With such a procedure, a density of vertical NWs < 0.1 NW/µm² was obtained. Such a low density allows single QD-NW spectroscopy on the as-grown sample instead of transferring the QD-NWs on a host substrate, thus the monolithic approach being conserved. Two steps were used to grow the QD-NWs (see methods). Briefly, a first step of axial growth of the InAs/InP QD-NWs was started by an initial InP stem growth during 12 min at 380 °C with an In BEP corresponding to an InP 2D layer growth rate equal to 1 µm/h and a V/III BEP ratio of 20. These growth parameters were chosen to ensure pure wurtzite (WZ) InP NWs.[18] A specific P/As and As/P switching procedure including the general shutter closing/opening[19] was then performed for the InAs QD growth to produce sharp InP/InAs direct and inverse interfaces (see methods). The growth of InAs was realized at 420°C for 3 sec with an In BEP corresponding to a 0.2 ML/s growth rate for a 2D layer, and a V/III BEP ratio of 20. This was followed by growing InP during 2 min with the same conditions for the stem. Then, a second step to realize the final axial/radial InP growth at a temperature chosen to lead to the targeted NW shape. All the NWs studied in this work were passivated by a 5 nm-thick wide band-gap $Ga_{0.15}In_{0.85}P$ shell, to prevent carrier surface recombination.

After the first step of axial growth, the QD-NWs are about 1.1 µm in length and 50-55 nm in diameter, with catalyst droplets in the 19-25 nm diameter range (after the sample cooling). Figure 1a shows the transmission electron microscopy (TEM) image of a typical InAs/InP QD-NW after this first growth step. The high-angle annular dark-field (HAADF) scanning transmission electron microscopy (STEM) image (in the [11-20] zone axis) in Figure 1b shows the InAs QD and reveals the InP/InAs interface abruptness, both direct and inverse. The InP NW and InAs QD exhibit a pure WZ structure thanks to the growth conditions used. Such a pure structure is highly important for QD based SPS because the presence of stacking faults or twins close to the QD will act as charge traps and could reduce the quality of the QD emission by the presence of space charge fluctuations.[31] The QD diameter D is in the 23-28 nm range and the height H is about 3-4 nm leading to an aspect ratio H/D ≈ 0.1. This aspect ratio was chosen because it is favorable to align the QD dipole perpendicularly to the NW axis[19,32,33] which is an essential condition to couple the emitted photons to the $HE_{11}$ mode of the NW.[26] Finally, it must be mentioned that, by comparing the InAs QD (or the Au droplet) diameter to the NW diameter, an InP radial growth of about 13 nm-thick takes place during this first step of axial growth.

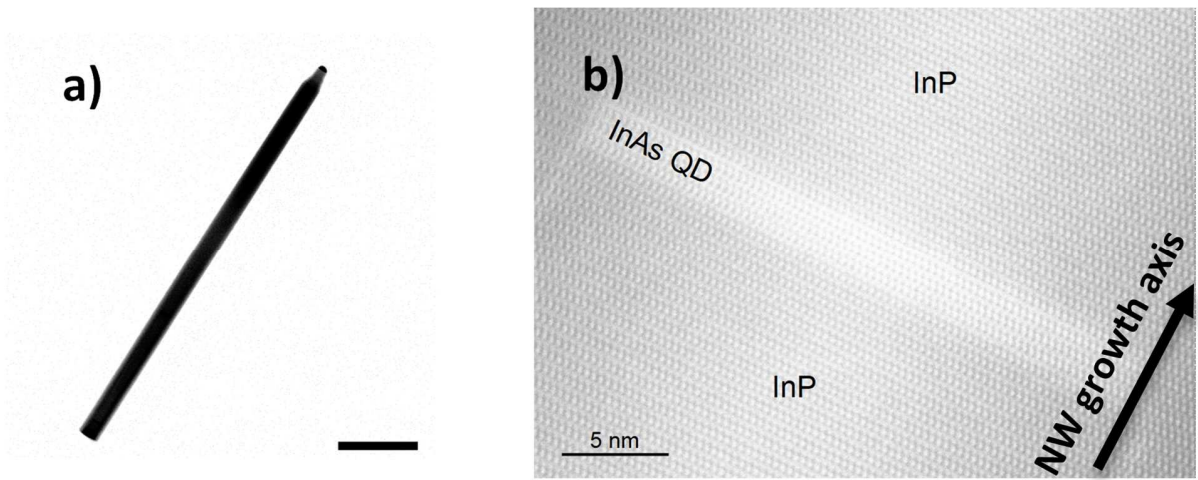

**Figure 1.** (a) TEM image of a typical InAs/InP QD-NW after the first step of axial growth (scale bar 200 nm). (b) HAADF-STEM image (in the [11-20] zone axis) of the InAs QD inside the InP NW (scale bar 5 nm).

Such NWs with a diameter in the 50-55 nm range are unable to waveguide efficiently the photons and the QD emission is inhibited by the too small NW diameter. It is therefore necessary to increase the diameter of the NWs in order to couple efficiently the QD emission to the fundamental NW guided mode $HE_{11}$ and thus avoid the QD emission inhibition.[34] Finite Difference Time Domain (FDTD) simulations were performed to determine the optimal NW diameter ($D_{NW}$) tailored to the QD emission wavelength ($\lambda$) using Opti-FDTD software (see methods).[35] The refractive index of the InP NW was chosen to be 3.2 corresponding to Zinc-Blende (ZB) InP. Figure 2a,b shows the designed QD-NW structure and the normalized spontaneous emission (NSE) of the InAs QD dipole as a function of $D_{NW}/\lambda$, respectively. In the case of a QD dipole oriented perpendicularly to the NW axis, an optimized QD NSE is reached for $D_{NW}/\lambda \approx 0.27$. This shows that a 350 nm NW diameter has to be targeted in the case of a QD emitting at 1.3 µm. In addition to this NW diameter condition, a second requirement concerns the upper part of the NWs that should present a small taper angle, α, to efficiently extract the guided photons from the NW tip into the free space as illustrated in Figure 2c.

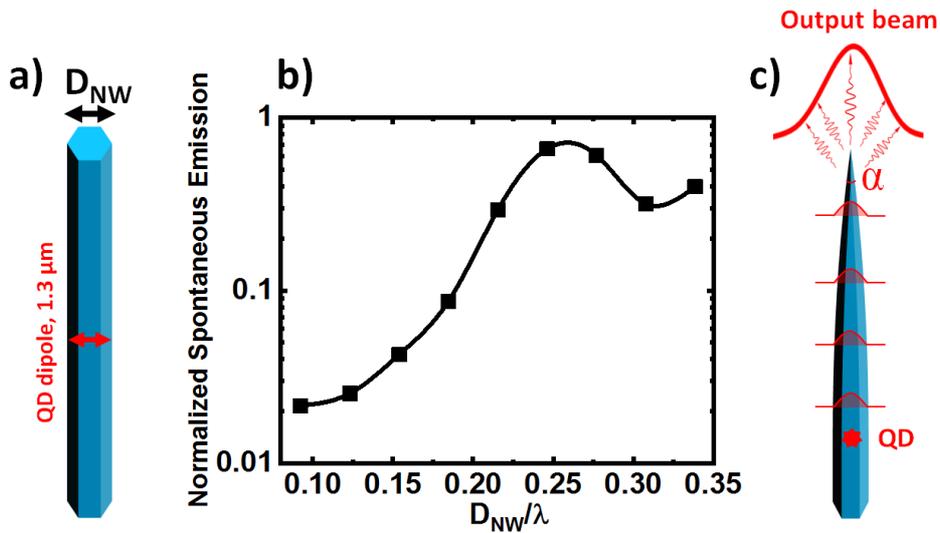

**Figure 2.** a) The designed QD-NW structure. The QD dipole is emitting at 1.3 µm and the NW diameter, $D_{NW}$, was varied. b) Normalized spontaneous emission of the QD dipole as a function of the $D_{NW}/\lambda$. c) Scheme of the tapered NW illustrating the effect of the taper angle, α, on the photon extraction from the NW waveguide.

A shell growth stage combining radial and axial growths was added to increase the NW diameter and to obtain a conical tapering of the NW tip. The later was carefully tuned by the choice of the InP shell growth temperature $T_G^{shell}$. Figure 3a-e shows the geometry of the NWs after the growth of InP shells for 50 min at 340°C, 360°C, 380°C, 400°C and 420°C. Figure 3f,g shows the evolutions of the NW length and diameter and of the taper angle α as a function of $T_G^{shell}$, respectively. With this procedure, the resulting taper angle can be varied from α ≈ 30° to a value as small as α ≈ 2°. In MBE, the In adatoms diffusing on the NW facets and reaching the catalyst droplet are the main source of In leading to the VLS growth.[36] For $T_G^{shell}$ equal to 340°C and 360°C (Fig. 3a,b), which are lower than the InP axial growth temperature (380°C), the radial growth occurs at the expense of the NW elongation rate due to the decrease of the In adatom diffusion length on the NW facets.[37] However, the NW axial growth is not completely suppressed during this step resulting in simultaneous radial and axial growths. Consequently, an intermediate region of conical morphology forms during this growth step resulting in a taper angle α ≈ 30° and 16° for $T_G^{shell}$ = 340°C and 360°C, respectively. On the contrary, if $T_G^{shell}$ is increased to 420°C (Fig. 3e), the axial growth is more favored than the radial one leading to a strong NW elongation. The NW length has reached an average value of 24 µm and the resulting taper angle is as small as α ≈ 2°. Note that these long NWs exhibit a pure WZ structure at the bottom part of the NW whereas few stacking faults and thin ZB insertions can be observed at the upper part of the NWs. Finally, by the introduction of the second growth step, the NWs can reach diameter in the 480-580 nm range which could be adapted to reach $D_{NW}$ = 350 nm as discussed earlier.

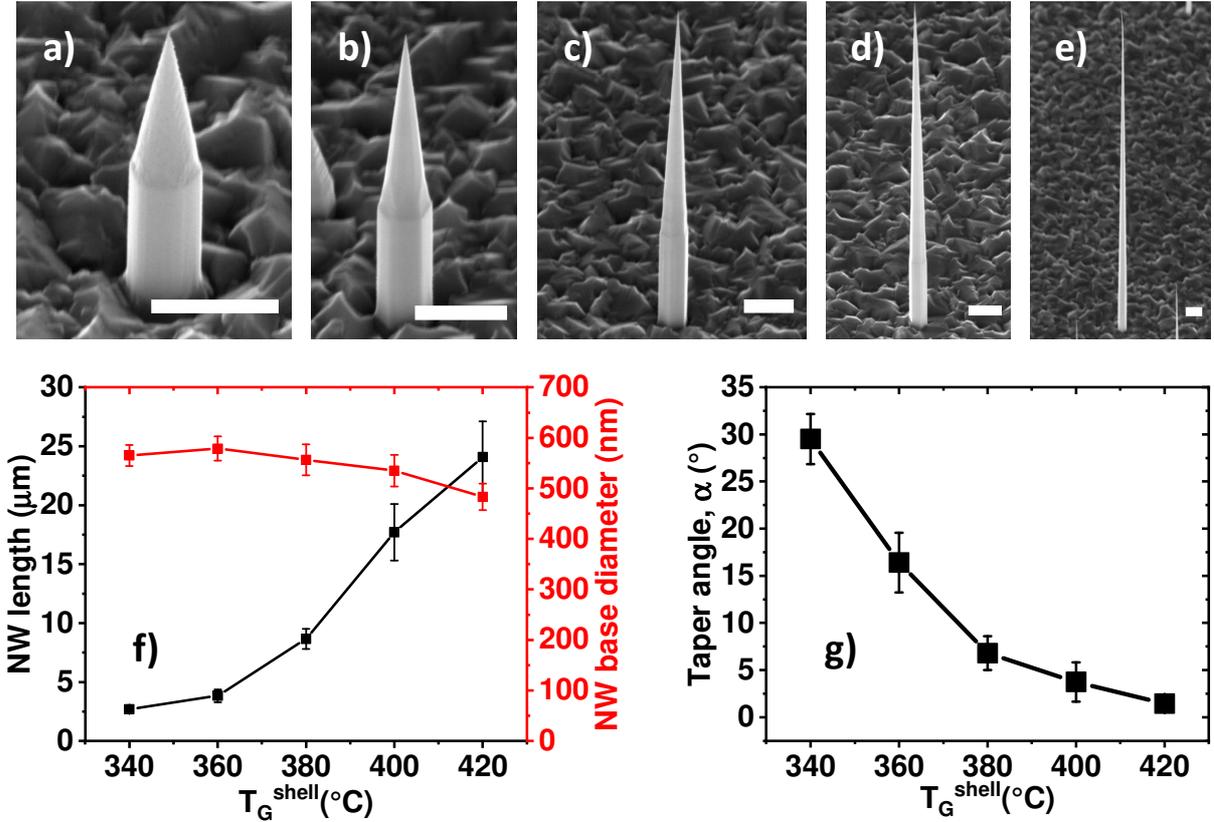

**Figure 3.** (a-e) SEM images (45° tilted) of single InAs/InP QD-NWs grown on Si(111) showing the NW shape depending on the shell growth temperature $T_G^{shell}$ at: a) 340°C: α ≈ 30°, b) 360°C: α ≈ 16°), c) 380°C: α ≈ 8°, d) 400°C: α ≈ 4°, e) 420°C: α ≈ 2° with the same shell growth time equal to 50 min (scale bars 1 µm). Evolutions with $T_G^{shell}$ of: (f) the NW length and diameter, and (g) the taper angle α.

Low temperature (12 K) photoluminescence (PL) highlights the effect of the taper angle on the QD emission. Figure 4a,b shows the emission spectra of the InAs QDs and the integrated PL intensity in the 0.85-1.05 eV energy range, respectively, for samples of NWs having different taper angle α. The PL spectra in Figure 4a were corrected by the QD-NW density to avoid any effect of the density on the QD emission intensity. As α decreases from 30° to 2°, the QD emission intensity increases gradually with a significant intensity when α becomes smaller than 15° and a 15.6-fold increase of the collected PL intensity is observed. FDTD simulations have been performed to understand this result.

First, the theoretical far-field emission profile of the five samples has been investigated using FDTD simulations (with the use of Rsoft FullWAVE software).[38] The calculation is performed using a 1.3 µm (0.95 eV) dipole oriented perpendicularly to the NW axis and positioned along the NW axis. The dipole is embedded in a cylindrical InP NW taking into account the equivalence between a circular NW of radius r and a hexagonal NW of side a≈1.1r as established in ref. 39. The numerical aperture of the PL setup is used to calculate the power radiated by the dipole and collected by the experimental setup (Fig. 4c). A 4-fold increase of the collected PL is expected when α is decreased.

Moreover, FDTD simulations have been performed to calculate the impact of the NW size, diameter and taper angle on the laser light absorption. We have calculated the total absorption in a cylindrical NW illuminated by an incident plane wave at 532 nm propagating along a direction parallel to the NW axis. The refractive index of ZB InP at 532 nm (n=3.7, k=0.43) is used and the size and geometry of the NWs are determined from the SEM images (Fig. 3). The total absorption as a function of the taper angle α is shown in Figure 4d revealing a 5-fold increase in the total absorption when α is decreased.

To conclude, if we assume that all the absorbed carriers are captured by the QD, a 20-fold increase of the QD emission can be expected when the taper angle α is decreased. This value is comparable to the experimental 15.6-fold increase. This difference is probably related to the non-perfect diffusion of the photogenerated carriers from the absorption area to the QD location. The comparison between the experimental results and the theoretical simulations (product of the normalized power by the normalized absorption) is shown in Figure 4b and confirms that the modification of the collected PL is both related to the modifications in the total absorption and in the far-field emission when α decreases.

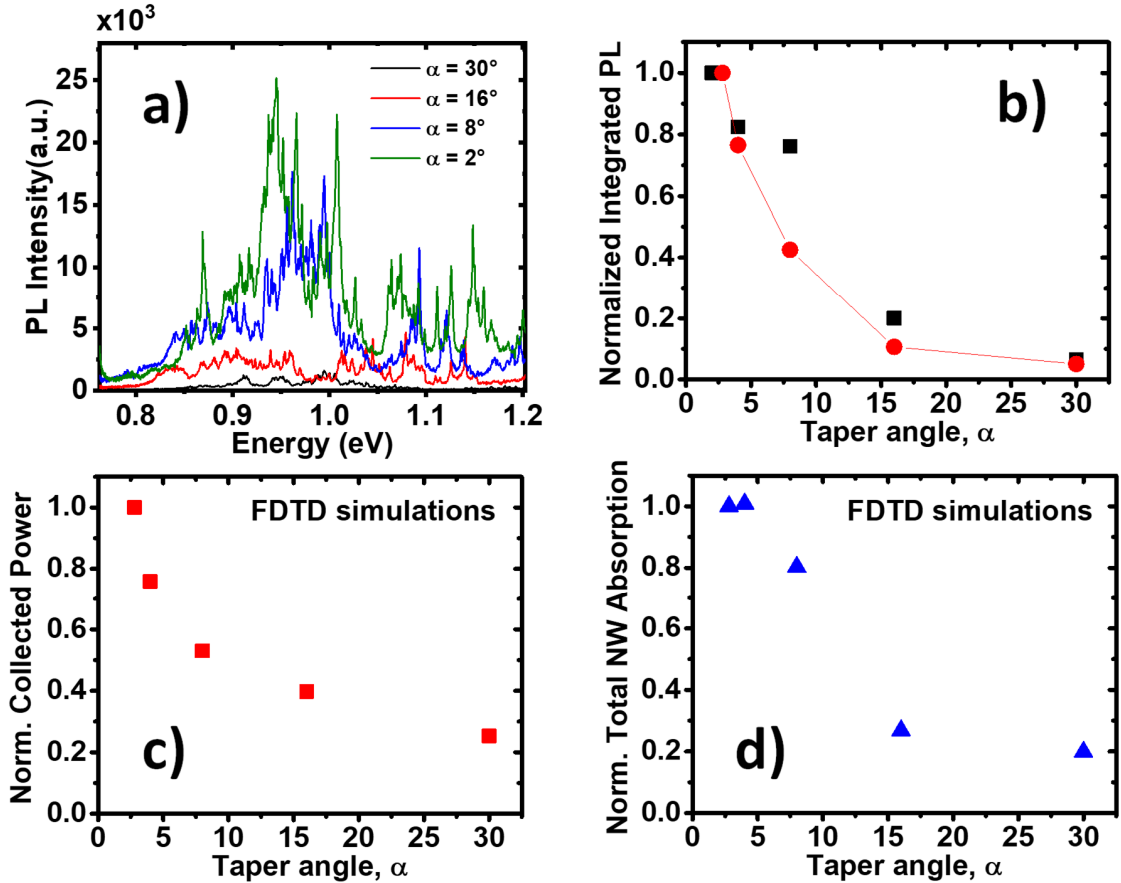

**Figure 4.** a) Low temperature (T = 12 K) PL showing the QD emission for samples of QD-NWs having different α. The PL intensity for each spectrum is normalized by the QD-NW density. b) Integrated PL intensity of the spectra in (a) in the 0.85-1.05 eV energy range, as a function of the taper angle α. The experimental results are the black squares. They are compared with numerical calculations coming from the FDTD simulations (red circles). c) Calculated power radiated by the dipole collected by the experimental setup. d) Calculated total absorption in the NW. The values in b), c) and d) are normalized using the NW with the smallest taper angle as a reference.

To obtain the optimal geometry with NW diameter ≈ 350 nm ($D_{NW}/\lambda \approx 0.27$), we have kept $T_G^{shell}$ = 420°C while the growth time was decreased from 50 min with $D_{NW}$ = 480±20 nm (Fig. 3e) to 35 min with $D_{NW}$ = 360±15 nm and 14 µm NW length (Fig. 5a). Contrary to the NW diameter and length, the NW taper angle α remained constant, 2°, since $T_G^{shell}$ was not modified. We have compared the QD-NW in Figure 5a. to another InAs/InP QD-NWs sample having a similar diameter $D_{NW}$ = 370±20 nm but shorter in length = 4±1.1 µm with wider α = 7°±2° (Fig. 5b). Such a QD-NW geometry was obtained by decreasing the $T_G^{shell}$ and the shell growth time to 380°C and 28 min, respectively.

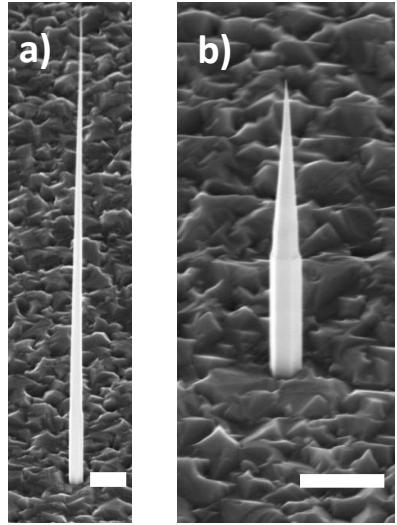

**Figure 5.** Tilted view (45°) SEM image of a typical InAs/InP QD-NW with an InP shell grown at: a) $T_G^{shell}$ = 420° during 35 min°, b) $T_G^{shell}$ = 380° during 28 min (scale bars 1 µm).

Micro-PL experiments were performed at 300 K and 12 K on the sample in Figure 5a to excite single QD-NWs on the as-grown sample. At 300 K, a single emission peak corresponding to the excitonic ground state transition is observed at low excitation power with an emission energy at 850 meV and a linewidth of 17 meV (Fig. 6a). At 12 K, micro-PL at low excitation power revealed a single QD line at 925 meV with a linewidth < 0.4 meV limited by the spectrometer resolution (Fig. 6b,c). The filling of the ground state and the first excited state transitions are also observed when the excitation power is increased. However, no clear exciton-biexciton behavior has been evidenced for these QD-NWs. This point could be explained by the fact that the InAs QD is located in the bottom part of a 14 µm-long NW. Due to the off-resonant laser excitation, carriers are photo-generated far away from the QD. As a consequence, carrier diffusion is required to explain the QD emission and it has been observed that this could favor charged (multi-)excitonic states in the QD.[40-43]

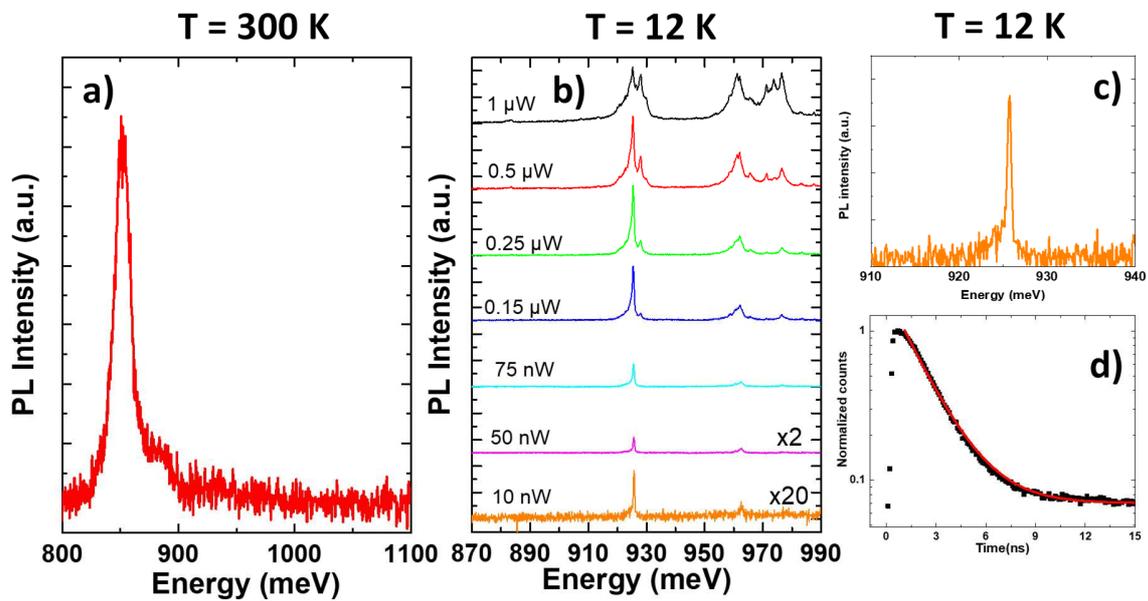

**Figure 6.** a) Micro-PL of a single InAs/InP QD-NW at 300 K, b) Excitation power dependent micro-PL at 12 K of a single InAs/InP QD-NW, c) Magnification of the InAs/InP QD-NW micro-PL in (b) at excitation power = 10 nW, d) Decay time measurement of a single InAs/InP QD-NW. Red curve: monoexponential fit.

The low temperature decay time measurements of a single InAs/InP QD-NW are presented in Figure 6d. The experimental lifetime $\tau = 1.8$ ns is similar to the theoretical values for ZB InAs/InP QDs[44] and to the experimental values obtained for ZB Stranski-Krastanov (SK) InAsP/InP QDs[45] for which values in the 1-2 ns range are usually reported. Our value is also comparable to the recently reported lifetime τ = 2.6 ns for InAsP QD in InP NWs emitting in the telecom band.[9] The measured lifetime is also consistent with the findings of Bleuse et al.[34] and Bulgarini et al.[46] where 0.5 to 2 ns QD lifetimes were registered when $D_{NW}/\lambda \approx 0.24$-$0.28$.

The effect of the NW taper angle α on the QD far-field emission was studied using Fourier spectroscopy at 300 K (see methods). We have compared the QD-NW in Figure 5a,b with α = 2° and α = 7°, respectively The InAs/InP QD-NW emission was collected with high numerical aperture (NA = 0.8) objective. The angular distribution of the photons emitted by the QD is represented by their k-vector in the reciprocal space with $k_x = (2\pi/\lambda)\sin(\theta)\cos(\phi)$ and $k_y = (2\pi/\lambda)\sin(\theta)\sin(\phi)$, where θ is the polar emission angle and φ is the azimuthal angle.

First, we have investigated using FDTD simulations (with the use of Rsoft FullWAVE software),[38] the theoretical far-field emission profile. We have designed a 1.4 µm wavelength dipole perpendicular to the NW axis embedded in a cylindrical InP NW, 320 nm in diameter, equivalent to a 360 nm diameter hexagonal NW[39] and studied the dipole's radiation pattern as function of the NW taper angle α. Figure 7a,b demonstrates the dipole radiation patterns for α = 2° and α = 7°, respectively. The simulations show that a Gaussian far-field profile is expected for both taper angles however the guided-mode profile for α = 2° expands such that the divergence of the output beam is reduced compared to that of α = 7° (Fig. 7c). The experimental far-field profile of a single QD emitting at ≈ 1.4 µm at room temperature is shown in Figure 7d,e for α = 2° and α = 7°, respectively. A good agreement between the experimental results and the FDTD simulations is observed where the QD's far-field emission profile for α = 2° exhibits minimum beam divergence. Differently, the emission profile of the QD in the NW with α = 7° is broader. This behavior is expected for higher α values: when the QD guided light reaches the NW taper, the modal effective refractive index of the guided light will experience a fast decrease as it travels along the taper reaching a value equal to that of air (n = 1). As a result, the QD light is emitted into free-space modes that do not correspond to the $HE_{11}$ mode.

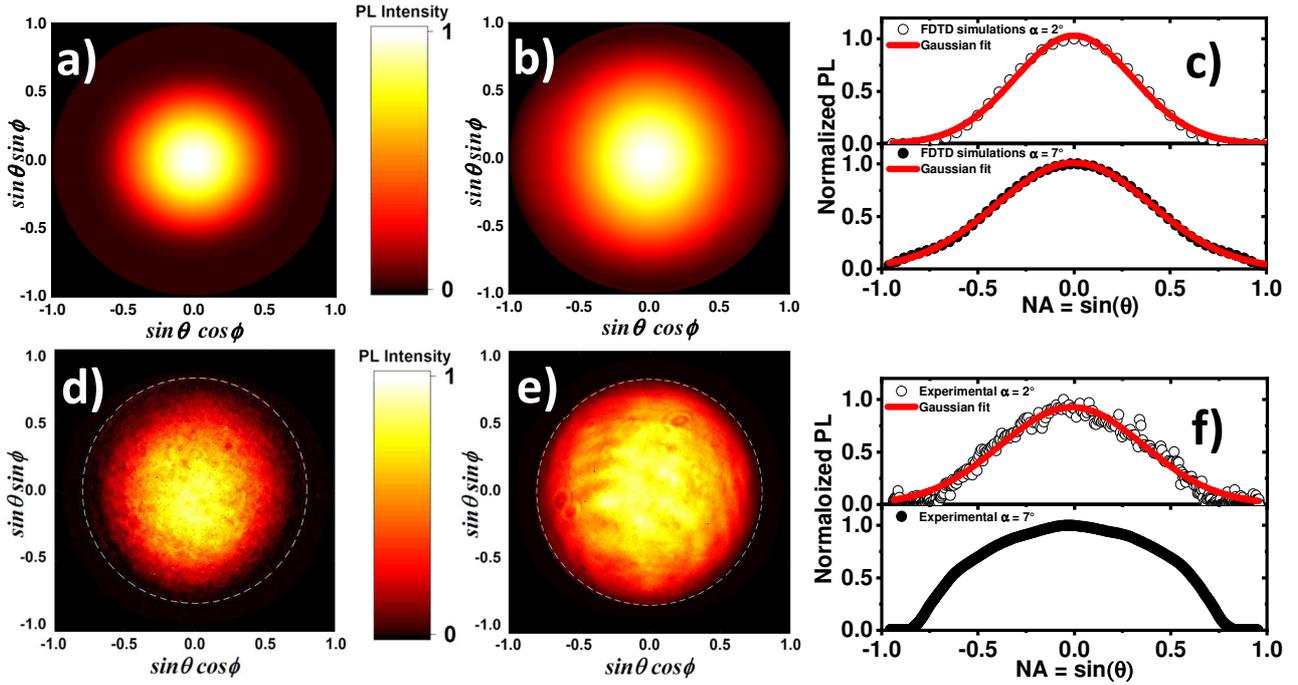

**Figure 7.** Normalized far-field emission profiles of a single InAs/InP QD-NW in K-space. FDTD simulations for a QD dipole perpendicular to NW axis emitting at 1.4 µm in a cylindrical tapered NW a) α = 2° b) α = 7°. c) 2D cut along the sinφ = 0 direction of the emission profile from FDTD simulations in (a) empty circles and in (b) filled circles. Experimental micro PL Fourier spectroscopy from a single InAs QD embedded in a tapered InP NW d) α = 2° e) α = 7°. The dashed circle corresponds to the maximum collection of the numerical aperture. f) 2D cut along the sinφ = 0 direction of the experimental emission profile in (d) empty circles and in (e) filled circles. Solid red lines are the Gaussian fits showing the Gaussian radiation patterns of the InAs QD. The emission intensity in all the Figures is normalized to the maximum PL intensity of each figure.

We have performed a 2D cut of the far-field profile for taper angles α = 2° and α = 7° (Fig 6f). The fit to the experimental data reveals the Gaussian pattern for α = 2° with a beam divergence angle θ = 30°±0.1°, however the broadening of the radiation pattern for α = 7° limits the possibility for fitting the data. This confirms that a small NW taper α < 7° is necessary to decrease the QD emission profile divergence and therefore improves the QD light coupling to optical fibers. Moreover, the Gaussian far-field profile of the QD is a clear evidence that the QD is mainly coupled to the fundamental, $HE_{11}$, mode of the NW.

The Overlap Integral (OI) between the 2D cut and the Gaussian fit was calculated using the following equation:

$$OI = \frac{\left(\int_{-\sin\theta_{NA}}^{\sin\theta_{NA}} I(\sin\theta) \times G(\sin\theta)\, d\sin\theta\right)^2}{\int_{-\sin\theta_{NA}}^{\sin\theta_{NA}} I(\sin\theta)^2 d\sin\theta \times \int_{-\sin\theta_{NA}}^{\sin\theta_{NA}} G(\sin\theta)^2\, d\sin\theta} \times 100\%$$

where $I(\sin\theta)$ is the intensity of the experimental 2D cut, $G(\sin\theta)$ is the intensity of the Gaussian fit (solid red line in Fig. 7f), and $\sin\theta_{NA}$ is the numerical aperture of the experimental setup (NA = 0.8). The fit of the experimental data with a Gaussian function shows a 98% overlap area. To estimate the collection efficiency of the objective, we assume a perfect Gaussian profile with a beam divergence of 30°. The transmission efficiency of the taper to an optics with collection angle $\theta_{NA}$ is:

$$T(\sin(\theta_{NA})) = \frac{\int_0^{2\pi} d\varphi \int_0^{\theta_{NA}} \frac{dI}{d\Omega}(\theta,\varphi)\sin\theta d\theta}{\int_0^{2\pi} d\varphi \int_0^{\frac{\pi}{2}} \frac{dI}{d\Omega}(\theta,\varphi)\sin\theta d\theta}$$

where $\frac{dI}{d\Omega}(\theta,\varphi)$ is the far-field profile described by a Gaussian pattern (solid red line in Fig. 7f). Our experimental data shows that a taper angle α = 2° is adequate to transmit 93% of the output beam into NA = 0.8. This is an evidence that nearly all the photons emitted from the NW top side are intercepted by the microscope objective with NA = 0.8. We can further calculate the source efficiency $\epsilon(\theta_{NA})$, defined as the probability of collecting a photon into the first lens of the optical set-up, with the following relation (adapted from ref. 24):

$$\epsilon(\theta_{NA}) = \frac{1}{2}\beta\frac{(1+|r_m|)^2}{(1+\beta|r_m|)}T(\sin\theta)$$

where $\beta = 0.9$ (from ref. 15) is the fraction of the QD spontaneous emission coupled to the $HE_{11}$ mode and $r_m$ is the modal reflectivity at the NW/substrate interface. In our case, we assume that the back-reflection is weak: $r_m = 0$. A source efficiency of $\epsilon(\theta_{NA}) \approx 42\%$ for NA = 0.8 is estimated which is similar to the state of the art of bottom-up NW based SPS emitting in the 850-950 nm range.[25] A modification of the current design is required in order to improve $r_m$ and be able to reach the record source efficiency of 72% obtained for a top-down photonic wires fabricated on a planar mirror.[15]

Finally, we have performed second-order correlation measurements (see Methods) at cryogenic temperature on a single InAs/InP QD-NW for the two different NW geometries in Figure 5a,b. Under continuous off-resonant laser excitation, coincidence counts were measured from an InAs QD emitting in the telecom O-band. Figure 8a,b shows the normalized corrected second-order correlation function $g^2(\tau)$ of a single QD emitting at 987 meV (NW with $\alpha = 2°$) and 933 meV (NW with $\alpha = 7°$), respectively, where coincidences involving background photons were subtracted. We have performed background suppression on the measured auto-correlation function, $g_u^2(\tau)$, to extract the auto-correlation function, $g^2(\tau)$, of a sole QD emission. Such correction is given by ref. 47: $g^2(\tau)-1 = (g_u^2(\tau)-1)/\rho^2$ where $\rho = S/(S+B)$ is the ratio of the signal, S, and background, B, photons. The antibunching dip below 0.5 at zero delay time is characteristic of a single photon emitter. The $g^2(0) = 0.32$ and $g^2(0) = 0.05$ are obtained from the theoretical fit to the data (solid red lines) for $\alpha = 2°$ and $\alpha = 7°$, respectively, taking into account the response function of the experimental setup.

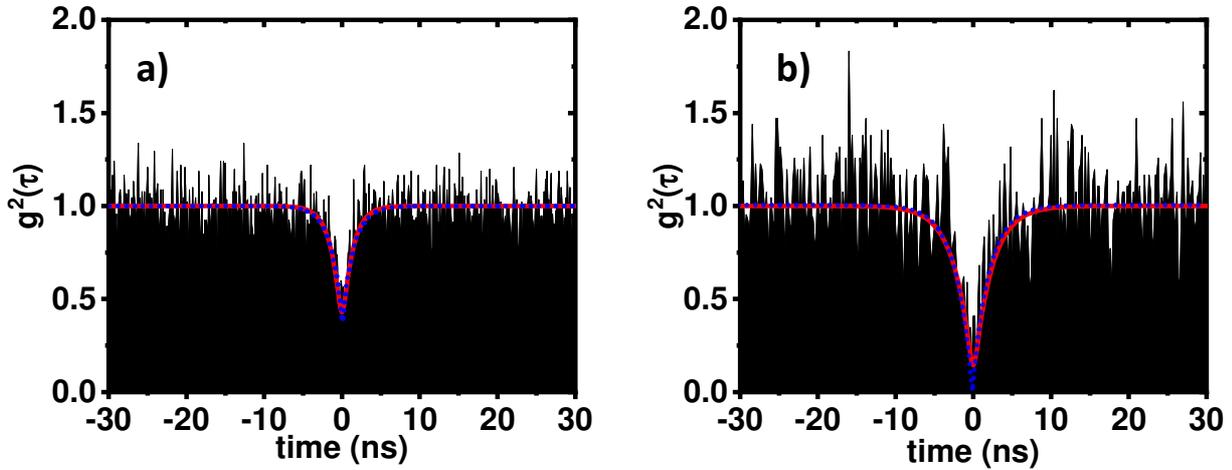

**Figure 8**. Second-order correlation measurements of a single InAs QD in an InP NW with: a) α = 2° emitting at 987 meV (1256 nm) resulting in a $g^2(0) = 0.32$ obtained from the fit (solid red line), b) α = 7° emitting at 933 meV (1328 nm) resulting in a $g^2(0) = 0.05$ obtained from the fit (solid red line). The blue dotted lines correspond to the same fit but without Gaussian convolution.

The high $g^2(0)$ value for the InAs/InP QD-NW with α = 2° could be explained by the fact that the NW length is 14 µm and the upper of the NW exhibits few stacking faults and thin ZB insertions resulting in carrier recombination, trapping and de-trapping mechanisms since the carriers are photogenerated at the NW upper part due to the off-resonant laser excitation.[48] Consequently, this could deteriorate the purity of the single photon emission from the InAs QD.[49] Improving the $g^2(0)$ value was achieved by removing the defects from NW upper part using lower $T_G^{shell}$ (380°C) Fig.5b. This $T_G^{shell}$ is similar to the axial growth temperature used to grow the first stem of the NWs resulting in a shorter defect free WZ QD-NWs from the bottom to the upper of the NWs compared to that of NWs grown with $T_G^{shell}$ = 420°C. Although the resulting NW taper angle α = 7° is relatively wide showing a broader far-field emission profile, the $g^2(0) = 0.05$ value is further ameliorated therefore decreasing the multi-photon emission probability. Thus, the balance between the NW taper angle and the NW crystal structure is highly important in order to have α as small as possible for a minimum output beam divergence while conserving the pure WZ structure of the NW for a pure single photon emission from the QD.[49] Notwithstanding that our $g^2(0)$ value is comparable to the recently published InAsP/InP QDs-NW telecom band SPS with $g^2(0) = 0.02$,[9] yet this is the first reported telecom band SPS based on InAs/InP QD-NWs monolithically grown on a Si substrate with a Gaussian far-field emission profile from the QD.

In conclusion, we have succeeded to monolithically grow a telecom band SPS based on a single InAs/InP QD-NW grown on Si(111). The NW geometry was controlled by using adapted shell growth temperature and growth time to reach a needlelike tapered NW with α = 2° and $D_{NW}$ = 360±15 nm. The far-field emission profile of a single InAs/InP QD-NWs with such a geometry allowed us to demonstrate a Gaussian radiation pattern thanks to the single mode waveguide supported by the NW and to the small taper angle α. This small α is highly efficient to extract more guided photons towards the collection optics with minimizing the divergence of the Gaussian radiation of the QD. However, growing NWs with a small α = 2° is accompanied with the formation of defects at the WZ NW upper part due to the elevated $T_G^{shell}$ (420°C) and presents a drawback for SPS with high $g^2(0) = 0.32$. For an optimal single photon generation, we have found a compromise between a NW taper angle α = 7° and pure WZ crystal structure along the total length of the QD-NW by using a lower $T_G^{shell}$ (380°C). Such a trade-off appeared to be highly important revealing a cleaner single photon emission from the QD with $g^2(0) = 0.05$. With respect to all these experimental results, this work presents an advanced and simple method for the growth and control of the InAs/InP QD-NW geometry serving as an original approach for the monolithic growth of a telecom band SPS on Si.

**Methods:**

**InAs/InP QD-NWs growth:** InAs/InP NWs were grown on Si(111) substrates using VLS assisted ss-MBE with Au-In droplets as catalyst. Before the growth, the Si substrate is exposed to UV light for 30 min followed by etching with a Buffered Oxide Etch (BOE) solution for 1 min and rinsed using de-ionized water for 1 min. Then, the substrate is bonded on the sample holder in air using indium glue at 260°C. During this procedure, a thin layer of $SiO_x \approx 1$ nm-thick is formed. Then, the sample is outgassed under ultra-high vacuum at 200°C for 15 min and introduced in the MBE reactor. Inside the MBE reactor, Au-In catalyst droplets are formed at 500°C by Au and In co-evaporating for 60 sec, with In and Au beam equivalent pressures equal to $2.5 \cdot 10^{-7}$ torr and $4.2 \cdot 10^{-9}$ torr, respectively, corresponding to an In/Au BEP ratio $\approx 60$. This step is followed by decreasing the sample temperature to 380°C corresponding to the optimal growth temperature to obtain pure WZ InP NWs, with an In flux corresponding to an InP 2D layer growth rate equal to 1 µm/h and a V/III BEP ratio = 20.[50] Prior to the NW growth, the $P_2$ flux is opened for 10 sec to form InP pedestals with an Au droplet on it.[19] Then, the In shutter is opened to start the InP NW growth. After 12 min, the growth is stopped for 2 min 20 sec by closing the general shutter (also the In shutter), while keeping the $P_2$ supply. During this growth interruption, the sample temperature is increased and stabilized (2 min) at 420°C which is favorable to minimize InAs radial growth during the InAs QD growth. Then, the $P_2$ flux is stopped and 10 sec after the $As_4$ flux is opened. 10 sec after, the InAs QDs are grown for 3 sec by opening the general shutter (also the In shutter) with an In BEP corresponding to a 0.2 ML/s growth rate for a 2D layer and an As/In BEP ratio = 20. After the growth of the InAs QD, the growth is stopped for 20 sec by closing the general shutter (also the In shutter) and stopping the $As_4$ flux. During this growth interruption, the sample temperature is decreased to 380°C. Then, 10 sec after the $P_2$ flux is opened and 10 sec after the general shutter (also the In shutter) is opened to grow InP for 2 min. The axial growth is stopped by closing the In shutter, while keeping the $P_2$ supply and the sample temperature is adjusted for the growth of the targeted InP shell. For this InP shell, the V/III BEP ratio = 20.

**FDTD simulations, designing a NW waveguide:** We have modelled a QD dipole emitting at 1.3 µm, located in the middle of a 1 µm long hexagonal InP NW with a refractive index n = 3.2 and oriented perpendicular to the NW axis. The QD dipole's normalized spontaneous emission (NSE) is calculated using the equation: NSE = $P_{NW}/P_{bulk}$, where $P_{NW}$ ($P_{bulk}$) is the classical radiation power for a dipole located in the NW (in bulk InP). The classical power radiated by the dipole is obtained by integrating the Poynting vector of that field along a surface that encloses the source.

**Photoluminescence and micro-Photoluminescence (µPL):** The samples are mounted inside a closed cycle helium cooled cryostat allowing measurements at T = 12 K. The QD emission is measured by a liquid nitrogen cooled Indium Gallium Arsenic (IGA) detector camera coupled to a monochromator. Continuous wave (cw) PL spectroscopy is carried out using a 532 nm diode-pumped solid-state laser source (spot size ≈ 200 µm). µPL is performed with a cw single mode fiber-coupled 780 nm laser source focused on the sample using a x50 microscope objective with a numerical aperture NA = 0.4 (spot size ≈ 4 µm).

**Angle-resolved photoluminescence with Fourier spectroscopy:** The sample is excited using a diode pump solid-state 532 nm laser at room temperature. The light is focused using a x50 microscope objective with a NA = 0.8 numerical aperture. The laser spot on the sample is ~ 2 µm in diameter allowing single QD-NW spectroscopy. The QD light emission is collected from the same microscope objective and its far-field is imaged on the sensor of an IGA camera with the use of a Fourier lens. This camera is also used in real space (i.e. without the Fourier lens) to align the laser spot with a single QD-NW. A long-pass filter with a cut-on wavelength of 1300 nm is used to remove the emissions from the InP core and shell and from the InP-InAs layer grown on the substrate.

**Time resolved and intensity correlation measurements:** The PL signal was detected by low-noise InGaAs photodiodes with a cut-off detection at 1700 nm, after a spectral selection with a long-pass filter at 1250 nm. The temporal decay was recorded by means of time-correlated-single-photon counting (TCSPC) measurements with an overall temporal resolution of 400 ps. Intensity correlation measurements were performed with a Hanbury Brown and Twiss (HBT) setup.

**AUTHOR INFORMATION**

**Corresponding Author:** *E-mail: ali.jaffal@insa-lyon.fr

**Funding Sources:** This work was financially supported by the network ULYSSES (ANR-15-CE24-0027-01) funded by the French ANR agency and the German DFG (PE 2508/1-1). The STEM studies were carried out on a microscope acquired as part of the TEMPOS project (ANR-10-EQPX-0050).

**Notes: The authors declare no competing financial interest.**

**ACKNOWLEDGMENT**

The authors thank the NanoLyon platform for access to equipments and J. B. Goure and C. Botella for technical assistance.